\input harvmac
\input epsf

\Title{\vbox{\rightline{EFI-97-47}\rightline{hep-th/9710217}}}
{\vbox{\centerline{Comments on Black Holes in Matrix Theory}}}
\vskip10pt
\centerline{Gary T. Horowitz}
\medskip

\baselineskip=12pt
\centerline{\sl Physics Department}
\centerline{\sl University of California}
\centerline{\sl Santa Barbara, CA 93106, USA}
\medskip
\centerline{\it and}
\medskip
\centerline{Emil J. Martinec} 
\medskip
\centerline{\sl Enrico Fermi Inst. and Dept. of Physics}
\centerline{\sl University of Chicago}
\centerline{\sl 5640 S. Ellis Ave., Chicago, IL 60637, USA}

\baselineskip=16pt
 
\vskip 2cm
\noindent

The recent suggestion that the entropy of Schwarzschild black holes 
can be computed in matrix theory using near-extremal D-brane
thermodynamics is examined.  It is found that the regime
in which this approach is valid actually describes 
black strings stretched across the longitudinal
direction, near the transition where black strings become
unstable to the formation of black holes.  It is argued
that the appropriate dynamics on the other (black hole) side of
the transition is that of the zero modes of
the corresponding super Yang-Mills theory.
A suggestive mean field theory argument is given for the entropy
of black holes in all dimensions. Consequences
of the analysis for matrix theory 
and the holographic principle are discussed.

\Date{10/97}

%
%
\def\journal#1&#2(#3){\unskip, \sl #1\ \bf #2 \rm(19#3) }
\def\andjournal#1&#2(#3){\sl #1~\bf #2 \rm (19#3) }

\def\ie{{\it i.e.}}

\def\cf{{\it c.f.}}

\def\etc{{\it etc.}}

\def\sst{\scriptscriptstyle}

\def\frac#1#2{{#1\over#2}}
\def\coeff#1#2{{\textstyle{#1\over #2}}}

\def\inbar{\,\vrule height1.5ex width.4pt depth0pt}
\def\IC{\relax\hbox{$\inbar\kern-.3em{\rm C}$}}
\def\IR{\relax{\rm I\kern-.18em R}}
\def\IP{\relax{\rm I\kern-.18em P}}

%
%
\def\np#1#2#3{Nucl. Phys. {\bf B#1} (#2) #3}
\def\npb#1#2#3{Nucl. Phys. {\bf B#1} (#2) #3}
\def\pl#1#2#3{Phys. Lett. {\bf #1B} (#2) #3}
\def\plb#1#2#3{Phys. Lett. {\bf #1B} (#2) #3}
\def\prl#1#2#3{Phys. Rev. Lett. {\bf #1} (#2) #3}

\def\prd#1#2#3{Phys. Rev. {\bf D#1} (#2) #3}

\def\jmp#1#2#3{J. Math. Phys. {\bf #1} (#2) #3}
\catcode`\@=11
\def\slash#1{\mathord{\mathpalette\c@ncel{#1}}}
\overfullrule=0pt

\def\LL{{\cal L}}
\def\MM{{\cal M}}

\def\PP{{\cal P}}

\def\lam{\lambda}

\def\underrel#1\over#2{\mathrel{\mathop{\kern\z@#1}\limits_{#2}}}

\catcode`\@=12


%

\def \sinh{{\rm sinh}}
\def \cosh{{\rm cosh}}

\def\bh{{\sst BH}}
\def\lc{{\sst LC}}
\def\D{{\sst D}}

\def\lpl{\ell_{{\rm pl}}}

%
\nref\BHreview{J. Maldacena, hep-th/9705078;
G. Horowitz, contribution to the Symposium
on Black Holes and Relativistic Stars
(dedicated to the memory of S. Chandrasekhar),
Chicago, IL, 14-15 Dec 1996; hep-th/9704072.}
\nref\joe{J. Polchinski, hep-th/9510017, \prl{75}{1995}{4724}; For a 
review see J. Polchinski, {\it TASI Lectures on
 D-Branes,} ITP preprint NSF-ITP-96-145, hep-th/9611050.
}
\nref\cbh{A. Strominger and C. Vafa, hep-th/9601029, \pl  {379}{1996}{99};
 C. Callan and J. Maldacena, hep-th/9602043, \np{472}{1996}{591};
G. Horowitz and A. Strominger, hep-th/9602051, \prl{77}{1996}{2368}.}
\nref\rotbh{J. Breckenridge, R. Myers, A. Peet and C. Vafa, hep-th/9602065,
\pl{391}{1997}{93};
J. Breckenridge, D. Lowe, R. Myers, A. Peet, A. Strominger 
and C. Vafa, hep-th/9603078, \pl {381}{1996}{423}.}
\nref\joegary{G. Horowitz and J. Polchinski, \prd{56}{1997}{2180};
hep-th/9612146.}
\nref\bfss{T. Banks, W. Fischler, S.H. Shenker and L. Susskind,
hep-th/9610043; \prd {55}{1997}{5112}.}
\nref\sussk{L. Susskind, hep-th/9704080.}
\nref\dira{M. Dine and A. Rajaraman, hep-th/9710174.}
\nref\dooo{M. Douglas and H. Ooguri, hep-th/9710178.}
\nref\sen{A. Sen, hep-th/9709220.}
\nref\seiberg{N. Seiberg, hep-th/9710009.}
%

%
%
\nref\dvv{R. Dijkgraaf, E. Verlinde, and H. Verlinde, hep-th/9704018.}
\nref\limart{M. Li and E. Martinec, hep-th/9704134.}
\nref\bfks{T. Banks, W. Fischler, I. Klebanov, and L. Susskind,
hep-th/9709091.}
\nref\ks{I. Klebanov and L. Susskind, hep-th/9709108; 
see also E. Halyo, hep-th/9709225.}
\nref\fhrs{W. Fischler, E. Halyo, A. Rajaraman, and L. Susskind,
hep-th/9703102.}
\nref\glf{R. Gregory and R. Laflamme, hep-th/9301052;
\prl{70}{1993}{2837}.}
\nref\gibbons{G. Gibbons, \npb{207}{1982}{337}.}
\nref\holo{G. `t Hooft, gr-qc/9310026; L. Susskind, hep-th/9409089,
\jmp{36}{1995}{6377}.}
%
\nref\bfk{T. Banks, W. Fischler, I. Klebanov, and L. Susskind,
hep-th/9711005.}
\nref\hoppe{J. Hoppe, hep-th/9702169.}
\nref\hoppenicolai{J. Hoppe and H. Nicolai, \plb{196}{1987}{451}.}
\nref\ghrw{J. Gauntlett, J. Harvey, M. Robinson,
and D. Waldram, hep-th/9305066; \npb{411}{1994}{461}.}
\nref\dkps{M. Douglas, D. Kabat, P. Pouliot, and S. Shenker,
hep-th/9608024; \npb{485}{1997}{85}.}
%

\newsec{Introduction}

String theory has recently provided a
statistical description of many aspects of black hole thermodynamics (for 
reviews, see \BHreview).
D-branes \joe\ have enabled a 
precise computation of the spectrum and interactions of a wide
variety of near-extremal charged \cbh\ and rotating \rotbh\ holes.
In addition, the black hole
correspondence principle \joegary\ gives a general relation (though not as
precise) between the entropy of essentially any black hole
and that of weakly coupled strings and D-branes.
However, in all cases, the  quantum states are described in weakly coupled
string theory, and the connection to large black holes is obtained
by increasing the string coupling and taking a low energy semiclassical limit.
In order to obtain a description of
the quantum states of a black hole directly in the black hole regime
one presumably needs a nonperturbative formulation of the theory.
Matrix theory \bfss\
purports to be such a formulation in the discrete light cone gauge \sussk.
In its eleven dimensional incarnation,  the quantum
mechanics of $N$ D-zerobranes with coupling $R$ appears to
reproduce eleven dimensional
supergravity\foot{Recent calculations \refs{\dira,\dooo}
indicate that there are subtleties in establishing this connection.},
with one direction compactified on a circle of radius $R$
and the total momentum $P=N/R$ . We will refer to $N$ as the number of
matrix `partons'. Compactifying further dimensions
leads to a super Yang-Mills (SYM) theory. The SYM hamiltonian
yields the invariant mass of the system via $H=M^2R/N$.
This prescription can be motivated \refs{\sen,\seiberg}
via a chain of dualities.
Given that the matrix theory prescription (in the dimensions where
it is known) is so closely tied to physics of D-branes 
in the large charge limit,
it is not surprising that the D-brane results 
on black holes can be carried
over quite successfully \refs{\dvv,\limart}, although the reinterpretation
of the calculations from this perspective is quite illuminating.
In particular, one is led to the idea \limart\ that the statistical
mechanics of (generalized) SYM theory could reproduce
black hole thermodynamics.

In two recent interesting papers, Banks, Fischler, Klebanov,
and Susskind \refs{\bfks,\ks} have argued
that the entropy of Schwarzschild black holes can be computed
from the SYM theory of near-extremal branes, at the special
point where the number of matrix partons $N$ equals the entropy $S$.
There are three main claims:

\item{1)} 
To fit a large black hole of size $r_0\gg R$ into the longitudinal box,
one must boost it so that its longitudinal size is contracted:
\eqn\xxmx{R>\Delta X_{\bh}\sim \frac{MR}{N} r_0}
\item{2)} 
Therefore $N>N_{min}\sim Mr_0\sim S$; $N\sim S$ is the minimal number of
matrix constituents required to describe the states of black holes.
At this lower bound, the black hole `just fits inside the box';
moreover, this value is `optimal' in the sense of the renormalization
group, in that increasing $N$ introduces a needlessly large number of
degrees of freedom, most of which are in their ground state.
\item{3)} One can apply SYM statistical mechanics to calculate 
the thermodynamic properties of the system at this threshold.
 For $D=8$ it was argued in \bfks\ that the threshold $N\sim S$
was just on the borderline of validity of this approach, since the 
effective temperature was so low that the typical wavelength was comparable
to the size of the system (in the appropriate holonomy sector corresponding
to multiply wrapped branes).

\medskip

A closer inspection yields a number of puzzles.
First of all, we will show that black holes do not contract longitudinally
when boosted.  If they did, it is likely that there would be a problem with
black hole thermodynamics. This is because the transverse size is presumably
invariant, so a Lorentz contraction would cause the horizon area to decrease.
But
this area is related to the black hole entropy which is a physical quantity
and should
not change under the boosts.  We will see that the horizon remains 
fixed and spherical 
with radius $r_0$ because the boost is effectively undone by 
the infinite gravitational redshift there.
How then can a large black hole fit inside the longitudinal box?
Secondly, the gluons and other excitations being counted
in the SYM thermodynamics carry field momentum, and in the
matrix dictionary (\cf\ \fhrs), the field momentum is
\eqn\ssms{ \PP_i=\int T_{0i}\propto \frac1{\Sigma_i}=\frac{RL_i}{\lpl^3}}
where $L_i$ are the lengths of the compact dimensions, $\Sigma_i$
are the lengths of the dual torus, and $\lpl$ is the eleven dimensional
Planck length. So these SYM momentum modes on the dual
torus
represent longitudinally stretched membranes in the original spacetime,
which are translationally invariant. They cannot represent a localized
object like a black hole.

We resolve these puzzles below.
While a boosted black hole does not contract, it does expand the geometry
near the horizon. In section 2 below,
we show that this effect is such that, precisely at $N=S$,
the geometry expands to the extent that the longitudinal box size
{\it at the horizon} is precisely $r_0$.  Thus it is not
the black hole that shrinks to fit in the box, rather
the box which expands to accommodate the black hole!
Moreover, it is clear that when the black hole
fills the box, it is on the verge of becoming a {\it black string}
stretched across the box.  Indeed, for zero momentum, it is known that
a black string becomes unstable when $R$ exceeds 
its Schwarzschild radius \glf.  We give
an entropic argument that the same phenomenon happens for the boosted
black holes and black strings.  The analysis of \refs{\bfks,\ks}
actually describes black strings, close to the transition point;
we will see that this corresponds to $N$ slightly smaller
than $S$.  As $N$ is increased, the temperature
of the SYM ensemble decreases, and for $N>S$ the system
is frozen into the dynamics on the space of zero modes.
Indeed, these zero modes on the dual torus represent objects which are not
longitudinally stretched in the original spacetime,
and are therefore the appropriate
degrees of freedom to describe black holes rather than black strings.

In section 3, we reconsider the transition point $N\sim S$,
approaching it from slightly larger values of $N$
where zero mode quantum mechanics should prevail.
A simple mean field theory analysis in the spirit of \bfss\
yields the black hole size $r_0$ and entropy $S$ as a function
of the mass $M$, {\it uniformly for any dimension} $D$.

This result leads us in section 4 to a reexamination of the 
`holographic principle' \holo.
In its mildest form (which we will
call the `weak holographic principle'),
 this principle states that the dynamics
of the theory depends only on data defined on a $D-2$ dimensional
spatial surface.  This may well be true in the matrix model,
given the intricate relation between the dynamics of the
transverse and longitudinal degrees of freedom embodied
in the matrices themselves.  The more virulent form of
the conjecture (which we shall call  the `strong holographic principle')
asserts that the physical size of objects increases with
boosting.  This idea is motivated by Bekenstein's proposal
that objects should respect a bound of one bit of information per
Planck area, and was in fact one of the prime motivations
for matrix theory.  It is not a necessary consequence
of the weak holographic principle, nor is it necessary
in order to account for black hole entropy in matrix
theory.
We construct examples of objects in matrix theory -- ensembles
of discretized membranes, to be specific -- which scale canonically
with longitudinal boosts, and therefore appear to
provide a counterexample to the claims of \bfss.

We conclude in section 5 with some speculations about how our results extend
to the regime $N\gg S$, and discuss some general consequences 
of our analysis for matrix theory and black holes.
Since our goal is to understand how matrix theory reproduces the familiar
scaling of  black hole entropy and size with mass,
we will often ignore numerical factors of order one,
such as the solid angle of spheres, \etc\


\newsec{Boosting black holes and black strings}

Let us begin by considering a D-dimensional spacetime
with one direction compactified on a circle of radius $R$ at infinity.
A Schwarzschild black hole in this spacetime corresponds to an infinite
periodic chain of D-dimensional black holes in the original
uncompactified space, and is not translationally invariant along the circle.
A black string is 
the product of a circle and a (D$-$1)-dimensional Schwarzschild black hole.
It is easy to show that for equal masses, the black hole has greater
entropy than the black string whenever $R$ is greater than the Schwarzschild
radius of the black hole $r_0$. 
In fact, the black string is known to be unstable in this
regime \glf. For $R < r_0$, the chain of black holes becomes
indistinguishable from the black string, which is now stable.

So far we have assumed that the momentum is zero. We now ask how things
change when we apply a boost along the circle\foot{The compactification
is different after the boost. Strictly speaking
one applies the boost to the uncompactified spacetime, and then identifies
points along the new spatial direction.}.
For the black string, one is boosting along  a symmetry
direction, so the metric is the same up to the usual coordinate
substitution $d\hat t\pm d\hat x_{D-1}=e^{\pm\beta}(dt\pm dx_{D-1})$.
The relevant piece of the metric transforms as
\eqn\strboost{
  -[1-(\rho_0/\rho)^{D-4}]d\hat t^2 + d\hat x_{D-1}^2 =
	-dt^2+dx_{D-1}^2 +(\rho_0/\rho)^{D-4}
		[\cosh\beta\;dt+\sinh\beta\;dx_{D-1}]^2
}
with the transverse part of the metric unchanged. 
If we periodically identify the coordinate $x_{D-1}$, the proper distance 
along this circle grows from $2\pi R$ at infinity to $2\pi R \cosh\beta$
at the horizon $\rho=\rho_0$. In a sense,
the momentum exerts a `pressure' on the geometry causing it to expand
near the horizon.

Now consider the D-dimensional black hole. If this were an ordinary
object, one might expect it to Lorentz contract when boosted. But
as mentioned in the introduction,
this would lead to problems with black hole thermodynamics. 
Fortunately, black holes
do not Lorentz contract. This follows from the fact that  every 
cross section of the event horizon has the same area. So in every reference
frame, the area of the horizon is the same. To see an explicit example,  
consider the four dimensional Schwarzschild metric in isotropic
coordinates
\eqn
\schwiso{ds^2 = - \left({r - r_0\over r+ r_0}\right)^2 d\hat t^2 +
\left (1+ {r_0\over
r}\right)^4 [d\hat z^2 + d\rho^2 + \rho^2 d\phi^2]}
where $r^2 = \hat z^2 + \rho^2$. These coordinates cover both asymptotically
flat regions, with the horizon  at $r=r_0$.
We now want to apply a boost
along the $\hat z$ direction. Since this is not a symmetry direction, there
is some ambiguity about how
one defines the boost in the interior of the spacetime.
A natural  choice is to simply set 
$d\hat t\pm d\hat z=e^{\pm\alpha}(dt\pm dz)$. The new horizon geometry
defined by $t=$ constant, $r=r_0$, is the surface in
\eqn\horgeo{ ds^2 = 16[\cosh^2\alpha dz^2 + d\rho^2 + \rho^2 d\phi^2]}
given by 
\eqn\defhor{ r_0^2 = r^2 = (\cosh\alpha \ z + \sinh \alpha \ t)^2
+\rho^2}
Differentiating this equation (with $t$ constant) yields
\eqn\horres{(\cosh\alpha dz)^2 = {\rho^2 d\rho^2 \over r_0^2 - \rho^2}}
so the induced metric on the horizon is 
\eqn\horans{ ds^2 = 16\left( {r_0^2 d\rho^2\over r_0^2 -\rho^2}+ \rho^2 d\phi^2
\right )}
which is completely independent of the boost and describes a round sphere.
So the event horizon does not Lorentz contract.
This is different from boosting a sphere in flat spacetime, since then
the $\cosh^2\alpha$ factor in \horgeo\ is absent. (It combines with a 
$-\sinh^2\alpha$
factor coming from $g_{\hat t \hat t}$. It is the infinite gravitational
redshift at the horizon which removes this cancelling term.)

Having established that the black hole does not Lorentz contract, we now
ask if the proper length of the circle expands near the horizon. It is 
clear from \horgeo\ that the answer is yes. In fact, at $t=0$ the horizon
is given by $r_0^2 = \cosh^2\alpha \ z^2 + \rho^2$, so in terms of the
coordinates $(z,\rho)$, there is an apparent
Lorentz contraction. It is the expansion of the metric near $r=r_0$ which
ensures that the horizon remains spherical. Thus
the net effect is similar to a real Lorentz contraction: One can fit large black
holes into small compactified spaces by boosting.

We now turn to eleven-dimensional supergravity
compactified to $D=11-d$ dimensions on a torus $T^d$ of volume $L^d$.
Consider a  black hole of mass $M=\lpl^{-9}L^d r_0^{D-3}$ which is given
a large boost in one of the $D$ directions which is then compactified on
a circle of radius $R$.
(The factor $\lpl^9/L^d\equiv G_\D$ is just the D-dimensional 
Newton constant.) 
The boosted black hole has energy and  momentum 
\eqn\momhole{\eqalign{
  E_{hole}=&~M\,\cosh\alpha=\lpl^{-9}{L^dr_0^{D-3}}\cosh\alpha\cr
  P_{hole}=&~M\,\sinh\alpha=\lpl^{-9}{L^dr_0^{D-3}}\sinh\alpha
}}
along the longitudinal direction.  
The boost does not change
the number of internal states of the hole, which remains
\eqn\enthole{
  S_{hole}\sim \lpl^{-9} L^d r_0^{D-2}\ .
}
Now choose the boost so that $P=S/R$.  This fixes $e^\alpha\sim r_0/R$,
and since the asymptotic longitudinal box size is $R$, at
the horizon the size is $r_0$ -- indeed the box just expands to accommodate
the black hole for this magic value of the boost.

Now let us compare this with the behavior of a boosted black string stretched
across the longitudinal direction.  Its energy, momentum,
and entropy are \gibbons
\eqn\strprops{\eqalign{
  E_{string}\sim &~\lpl^{-9}L^dR\rho_0^{D-4}[a+\cosh\,2\beta]\cr
  P_{string}\sim &~\lpl^{-9}L^dR\rho_0^{D-4}\sinh\,2\beta\cr
  S_{string}\sim &~\lpl^{-9}L^dR\rho_0^{D-3}\cosh\,\beta\ ,
}}
where $a$ is a constant of order one. The black string will be stable
provided the length of the horizon is less than its Schwarzschild radius
$\rho_0$. Since this length increases by  $e^\beta$ (for large $\beta$)
as we boost, the instability begins when $e^\beta \sim \rho_0/R$. This implies
$P\sim S/R$.
So we see from both black hole and black string considerations, that
the condition $N=S$ (where $P= N/R$)
marks the transition between these two configurations. Clearly, when
$P=S/R$, if the black hole and black string have the same momentum, then
they have the same entropy as well.

One might be puzzled by the different scaling of the energy and momentum
in \momhole\ and \strprops\ under a boost. The difference arises due to
what is implicitly held fixed. In \momhole, one starts with 
an infinite chain of black holes of mass $M$
in the spacetime with the longitudinal
direction uncompactified.  If the initial separation (asymptotically)
is $R$, after the
boost, the  energy of each black hole is $M\cosh\alpha$ and the
new separation  will be Lorentz contracted
$R/\cosh \alpha$. If we insist that the separation after the boost is $R$,
there are two options. One can periodically identify after
$O(\cosh\alpha)$ black holes are included. This produces another factor of
$\cosh \alpha$ in \momhole\ and \enthole\ so that the  energy, momentum,
and entropy now scale like 
the black string \strprops. Alternatively, one can increase the initial
separation between the black holes to $R\cosh\alpha$,  which insures that the
separation after the boost will be $R$. This was implicitly assumed in
\momhole\ and \enthole. 

In terms of matrix theory, the natural variables to hold fixed are $R$
and the invariant  mass $M$.\foot{We assume that $R$ is 
the radius of a spacelike circle, as in \bfss. However, since we are always
in the regime of very large boosts, the distinction (in this
Lorentz frame) between spacelike and lightlike compactification
in the longitudinal direction is not expected to be important.}
The light cone energy is $E_{\lc}=E-P$
and $P= N/R$,
so $M^2 = E_\lc N/R$.
Eliminating the parameters $r_0$, $\alpha$,
$\rho_0$, and $\beta$ in favor of $P= N/R$ and 
$E_{\lc}=E-P$ yields
\eqn\compare{\eqalign{
  S_{hole}\sim &~(\lpl^9/L^d)^{\frac1{D-3}}
	M^{\frac{D-2}{D-3}}\cr
  S_{string}\sim &~N^{1/2}\;(\lpl^9/L^d)^{\frac1{D-4}}
	\left(\frac {E_\lc}R\right)^{\frac{D-2}{2(D-4)}}\cr
	\sim &~N^{-\frac{1}{D-4}}\;(\lpl^9/L^d)^{\frac1{D-4}}
	M^{\frac{D-2}{D-4}}\ . 
}}
If $S_{hole}=S_{string}$ for some choice of mass $M$ and boost $N$, then
clearly increasing $N$  causes the black hole to have greater entropy
and decreasing $N$ causes the black string to have greater entropy.

This behavior is perfectly compatible with our expectations
from SYM statistical mechanics.  There, for fixed $R$ and $M$,
$E_\lc=M^2R/N$ decreases as $1/N$, \ie\ the ensemble becomes
colder with increasing $N$. At high temperature, corresponding to small $N$,
the SYM theory is roughly a
gas of interacting supergluons. As mentioned in the introduction, these
modes correspond to longitudinal membranes (and fivebranes if $d\ge4$)
in the original spacetime, and thus should describe states of a black string.
This is the situation described for instance in \limart.
As we increase $N$, this description remains valid until
the temperature of the
gas drops to the point that the thermal wavelength is comparable to 
the effective size of the dual torus on which the SYM
is defined.  Then
the gluonic degrees of freedom freeze out, leaving quantum
mechanics on the space of zero modes.  The statistical mechanics
is that of this quantum mechanical system (including the
various global fluxes on the internal torus).  These states
describe objects which are not longitudinally wrapped in the original spacetime,
and which can thus be localized in the longitudinal direction.
This system describes black hole states. From the above analysis, we see that
the transition occurs when $N\sim S$.

The calculations of \refs{\bfks,\ks} attempt to explain the black hole entropy
by approaching the black hole -- black string transition from the `wrong' 
side, using
the equality of  the density of states at the transition to
infer the entropy on the other side.  The procedure is 
similar in spirit to the black hole correspondence principle,
in which one infers the black hole spectrum from the
string spectrum by matching their densities of states
at the transition from one to the other, and using the
known string spectrum.  Here however, one is on much 
shakier ground; much less is known about how to compute
the SYM entropies from first principles
(although the case of 3+1 SYM corresponding  to $ D=8$ is on a somewhat firmer
footing).


\newsec{A direct approach to black hole entropy}

One might hope to 
arrive at the black hole entropy more directly,
by an analysis of the zero-mode quantum mechanics 
that begins to dominate just above the transition.
Indeed, a mean field analysis \bfss\ appears to capture the
essential physics.  When the matrix partons are sufficiently
far apart, the `fast' off-diagonal matrix element dynamics 
can be integrated out.  Treating the partons in mean-field
approximation, the one-loop effective Lagrangian 
for the zero modes (of matrix theory compactified on $T^d$, with $D=11-d$) 
has the structure
\eqn\oneloop{
  \LL_{{\rm eff}}=\frac{Nv^2}{R} + \frac{N^2\lpl^9v^4}{R^3L^d\,r^{D-4}}\ .
}
The parton mass is $1/R$ due to the origin of matrix
theory in ten-dimensional D-zerobrane physics.  
Recall that the factor $\lpl^9/L^d\equiv G_\D$ is just the D-dimensional 
Newton constant, so the second term can be interpreted as the gravitational
self-energy of the partons due to their relative motion.

The dynamics determined by this Lagrangian is rather complicated, but
mean field theory arguments indicate that there are solutions where
the partons remain in a bounded region of space of radius $r_0$
for an extended period of time. 
The virial theorem then tells us that the two terms in the
effective Lagrangian are of the same order.
We will assume that the partons saturate the uncertainty bound,
\eqn\heisenberg{
 {r_0 v\over R}\sim 1\ .
}
(Since \oneloop\ is derived under the assumption of nonrelativistic motion,
$v \ll 1$, the size of the bound
state must be much larger than the longitudinal box size
as measured at infinity.)
These assumptions determine a relation
between $N$ and the size of the bound state:
\eqn\Nsize{
  N\sim (\lpl^{-9}L^d)r_0^{D-2}\ .
}
The typical energy scale is then
\eqn\Emf{
  E_\lc=\frac{M^2R}{N}\sim (\lpl^{-9}L^dR)r_0^{D-4}\ ,
}
leading to a typical size of the bound state in terms
of the mass:
\eqn\msize{
  M\sim (\lpl^{-9}L^d) r_0^{D-3}\ .
}
Since $\lpl^{-9}L^d=1/G_\D$,
we recognize the relation between
the mass and Schwarzschild radius of a black hole!
Now consider the mass-entropy relation for D-dimensional
black holes; using \Nsize, \msize, we have
\eqn\Smf{
  S\sim (\lpl^{-9}L^d)^{\frac1{D-3}} M^{\frac{D-2}{D-3}}
   \sim \lpl^{-9}L^d r_0^{D-2}\sim N\ .
}
This is already clear from \Nsize\ -- the number of partons
is the surface area of the bound state in Planck units.
In other words, the black hole entropy is the number of partons
up to coefficients of order unity.
One can argue that the entropy in the partons is also of order
$N$ if they are effectively distinguishable\foot{Recently, 
the importance of using Boltzmann statistics was stressed in \bfk.},
since each parton has several polarization states.
Notice that this argument 
works uniformly in all dimensions $D$, and does not require
independent conjectures about the SYM thermodynamics.
The basic assumptions are simply (1) mean field theory \oneloop\ is applicable;
and (2) the system is in a minimal uncertainty bound state. 

As we saw in the last section, since $S\sim N$, one is again at the
transition between black holes and black strings. However since
the above analysis only concerns the quantum mechanics
of the zero modes, it approaches the black hole/black string transition
from the black hole side, rather than the black string
side as in \refs{\bfks,\ks}.  


\newsec{Zero mode dynamics and the holographic principle}

What can we expect for the zero mode quantum mechanics
as we increase $N$ to move away from the transition point?
The answer depends crucially on what happens to the bound state's
characteristics, in particular its transverse size, as we boost it
by increasing $N$.  The strong holographic principle would
predict that the transverse size $r$ increases with $N$,
leaving us with two scales -- the holographic size $r_h$
and the Schwarzschild radius $r_0$.  This seems awkward, since slow
scattering experiments 
will presumably depend only on $r_0$ and not on $r_h$.

A set of classical solutions studied by Hoppe \hoppe\
reveals a more canonical boost behavior, at least at the classical
level.  These solutions are matrix discretizations of those
found in \hoppenicolai.  Consider the ansatz
\eqn\ansatz{
  X^i(t)= x(t)r^i_{\ j}(t)\MM^j\ ,
}
with $x(t)$ an overall pulsation; 
$r^i_{\ j}(t)=exp[\varphi(t) \Omega]$ a rotation 
of constant angular momentum ${\rm L}=R^{-1}x^2(t)\dot\varphi(t)$;
and $\MM^j$ a fixed matrix
\eqn\Mansatz{
  \vec \MM=\coeff1{2\sqrt2}\Bigl(U+U^{-1},-i(U-U^{-1}),V+V^{-1},-i(V-V^{-1}),
	0,\ldots,0\Bigr)
}
in terms of the 't Hooft matrices $U$, $V$, satisfying
$UV=\omega VU$, $\omega=exp[2\pi i/N]$.  One may take the rotation
to have $\vec \MM$ as an eigenvector, $\Omega^2\vec \MM=-\mu \vec \MM$.
The algebra of the 't Hooft matrices gives
\eqn\Malg{
  \sum_{j}[[\MM_i,\MM_j],\MM_j]=\lambda \MM_i\quad,\qquad\lambda=2\sin^2(\pi/N)\ ,
}
and the solution to the classical equations of motion of the matrices $X^i$
boils down to that of the overall pulsation
\eqn\pulse{
  \frac{\ddot x}{R}+\lambda R x^3-\frac{\mu R{\rm L}^2}{x^3}=0\ .
}
(In this section we measure $x$ in Planck units.)
At large $N$, one has $\lam\sim N^{-2}$, 
and the conserved energy
is simply
\eqn\Econs{
  E_\lc\sim  N\Bigl[\frac{\dot x^2}{R} + \frac R{N^2} x^4 
			+ \frac{\mu R{\rm L}^2}{x^2}\Bigr]\ .
}
Since $E_\lc \propto 1/N$, the 
relevant scales are $x\sim 1$, 
$t\sim N/R$, 
and ${\rm L}\sim 1/N$.
In other words, the transverse size remains constant, and the motion
slows down as the system is boosted -- canonical
boost behavior.  Of course, the true test of the system
vis \`a vis the strong holographic principle is what happens
when quantum fluctuations are turned on.  Naively, these
ought simply to lead to the gravitational interactions between
the various bits of membrane.
There will, of course, also be zero-point fluctuations which grow without
bound as the cutoff $N$ is removed; however, these do not usually
affect the size of objects as seen in scattering experiments with slow
heavy probes.
For example, macroscopic strings appear local down to the string scale \ghrw,
and D-branes in slow relative motion may be localized down to 
the Planck scale and beyond \dkps. In the large $N$ limit, the quantum
membrane considered here should resemble a D-twobrane.

A membrane is not a black hole, of course; from \Econs,
its size scales as $M^2 R/N\sim x^4R/N$; the mass is quadratic
in the radius in any dimension (as opposed to $M_\bh \sim r_0^{D-3}$)
simply because the mass is the membrane area in Planck units.
However, one could imagine assembling a black hole from
a sufficient number of little nuggets of membrane, say $k$
of them, each looking semiclassically like \ansatz, and collapsing in
on one another under gravitational attraction.  
In other words, the full $kN\times kN$ matrix $X^i$ would decompose into
blocks of size $N\times N$ of the form \ansatz; the gravitational
attraction between different blocks comes from integrating out
the off-diagonal blocks.  The entire system appears to
obey canonical scaling under boosts $N\rightarrow e^\alpha N$.


\newsec{Discussion and Speculations}

Let us now consider what happens to the black hole states in matrix theory
in the limit $N\gg S$. The following remarks will necessarily be rather
speculative, since reliable calculations are not yet available in this
regime.
If the transverse size remains constant under boosts,
as suggested by the preceding analysis, then the partons become denser
as $N$ increases. 
It seems likely that strongly interacting clusters will form.
Within each cluster, the Born-Oppenheimer approximation 
will no longer be valid. This is because a given matrix parton is
close enough to the other partons in the cluster so that the nonabelian
degrees of freedom can
no longer be consistently integrated out.
The coherent interaction within a cluster should be more 
`membrane-like' than `graviton-like', since the commutator
term in the matrix Hamiltonian {\it is} the membrane area element.
The interaction between clusters 
might still be treatable in the Born-Oppenheimer approximation.

The typical size of a cluster can be estimated using Hawking radiation.
In the rest frame of the black hole, the Hawking radiation has
characteristic wavelength
of order the Schwarzschild radius $r_0$.
A boost to the transition point $P=S/R$ is such that
the longitudinal component of this radiation is Lorentz contracted to 
the box size $R$ \bfks.  An additional boost to $P= N/R$ $(N\gg S)$
will make the characteristic longitudinal momentum of a 
Hawking quantum $p_{||}=N/SR$. In matrix theory, this corresponds to
a (threshold) bound state of $N/S$ partons.  
Partons in the black hole must therefore be strongly correlated
over domains containing approximately this many partons.
This observation leads one to expect that there will be 
roughly $S$ clusters, each with approximately $N/S$ partons.

For a fixed mass black hole, the energy $E_{LC} = M^2 R/N$ decreases
as $N$ increases and the system becomes colder. Since the partons are
becoming denser, colder, and more strongly interacting, one can think
of this phase as a `parton liquid' 
(in contrast to the `gas' phase of the Born-Oppenheimer approximation
that governs well-separated partons).
If the transverse size remains constant under boosts, as suggested above,
the typical virial velocities decrease as $1/N$. For large $N$, this 
appears to violate the uncertainty bound \heisenberg. However, the bound on the
velocity of a cluster of $N/S$
partons is decreased by $S/N$ (since the mass is larger), so there is no
contradiction.
Of course, individual partons' velocities cannot violate the uncertainty 
principle. In order not to contribute too much to the 
overall energy, the partons within each cluster 
or domain
must be very nearly in their ground state (so their energy
approximately cancels between bosons and fermions due to supersymmetry).
This is in accord with 
the assertion of \bfks, that most of the partons must be
in their ground state for $N\gg S$.

The total energy $E_\lc$ will be distributed among kinetic energy of the
clusters,
gravitational potential energy (after integrating out
fast nonabelian modes), and `membrane stretching energy'
from the slow nonabelian modes of nearby (and strongly correlated) partons.
Computing the properties of the black hole in this regime will
require understanding how the system apportions its energy budget
among these, and perhaps other, aspects of the dynamics.

We have argued that matrix theory can describe some essential properties
of Schwarzschild black holes. This may seem surprising in light
of recent indications that matrix theory has difficulty reproducing
eleven dimensional supergravity \refs{\dira,\dooo}. 
We believe that the black hole
results indicate that matrix theory does capture the essential degrees of
freedom of the theory. It is possible that some detailed aspects 
of the matrix dynamics
may need to be modified, but the gross features are not likely to be affected.

We believe our analysis contains other lessons about
matrix theory as well.  It shows that the localized states 
of matrix theory are encoded in the zero mode dynamics of the
generalized SYM theory that defines matrix theory in a particular
compactification.  All the nonzero modes which are
the source of ultraviolet difficulties in the quantum theory
(and an apparent stumbling block in defining the theory
in compactifications to low dimensions) are longitudinally
stretched objects which must decouple in the
limit of interest $N,R\rightarrow\infty$.
Since the great success of matrix theory seems to stem from
its ability to exhibit all the dualities of M-theory,
one might wonder whether there is a truncation
of the dynamics to the zero mode sector which respects
the dualities while throwing away all the troublesome
aspects of the nonzero mode dynamics.  
What would change with dimension would be the combinatorics
of the bound states of various fluxes.

The zero mode sector of matrix theory at finite $N$ is
(so far) a theory of the electrically charged objects of M-theory,
containing a finite number of gravitons and/or discretized
membranes.  A complete theory must include the fluxes
corresponding to the magnetic objects as well -- the fivebrane
and six-brane (Kaluza-Klein `monopole').  Since these objects
are solitonic in nature, it is unlikely that they will be present
in the finite $N$ theory; rather, they are
`condensates' of a nonperturbative number of partons.
This may explain the difficulties encountered to date with
matrix theory on $T^5$ with transverse fivebranes,
and the apparent lack of a candidate for the theory 
at finite $N$ on $T^6$ and beyond.
While the scaling analysis of section 3 seems to work
in any dimension, it is likely that a proper understanding 
(especially for $N\gg S$) will have to incorporate
these fluxes in the dynamics for $D\le 6$, where the
continuum $N\rightarrow \infty$ theory may be needed.

Even without an understanding of such magnetic fluxes, there
remains a fascinating condensed matter problem to determine the thermodynamics
of the `parton liquid' whose properties appear to govern
black hole thermodynamics in quantum gravity.
Reproducing  properties of black holes when $N\gg S$
will teach us a great deal about the Lorentz covariance properties
of matrix theory, and should be somewhat simpler than the threshold bound
state problem for gravitons.


\vskip 1cm
\noindent{\bf Acknowledgments:} 
We are grateful to 
Miao Li,
Samir Mathur, and Philippe Pouliot
for discussions.
This work was supported by DOE grant DE-FG02-90ER-40560, and
NSF grant PHY95-07065.

\listrefs
\end